\documentclass{aa}  

\usepackage{graphicx}
\usepackage{txfonts}
\usepackage{multirow}
\usepackage{color}
\usepackage{makecell}
\usepackage[colorlinks=true,allcolors=blue]{hyperref}
\begin{document}

   \title{Investigation of the subsurface structure of a sunspot based on the spatial distribution of oscillation centers inferred from umbral flashes}

%   \subtitle{I. Overviewing the $\kappa$-mechanism}

   \author{Kyuhyoun Cho
          \inst{1}, 
          Jongchul Chae\inst{1},
          \and
          Maria S. Madjarska\inst{2, 1}
          }

   \institute{Astronomy Program, Department of Physics and Astronomy, Seoul National University, 
   		   	Seoul 08826, Republic of Korea\\
              \email{chokh@astro.snu.ac.kr}
         \and
             Max Planck Institute for Solar System Research, Justus-von-Liebig-Weg 3, 37077, G\"ottingen, Germany\\
             }

%   \date{Received September 15, 1996; accepted March 16, 1997}

% \abstract{}{}{}{}{} 
% 5 {} token are mandatory
 
  \abstract
{The subsurface structure of a solar sunspot is important in the stability of the sunspots and the energy transport therein. Two subsurface structure models have been proposed, the monolithic and cluster models, but no clear observational evidence supporting a particular model has been found so far. To obtain clues about the subsurface structure of sunspots, we analyzed umbral flashes in merging sunspots registered by IRIS Mg {\sc ii} 2796 \AA\ slit-jaw images (SJIs). Umbral flashes are regarded as an observational manifestation of magnetohydrodynamic (MHD) shock waves originating from convection cells below the photosphere. By tracking the motion of individual umbral flashes, we determined the position of the convection cells that are the oscillation centers located below the umbra. We found that the oscillation centers are preferentially located at dark nuclei in the umbral cores rather than in bright regions such as light bridges or umbral dots. Moreover, the oscillation centers tend to deviate from the convergent interface of the merging sunspots where vigorous convection is expected to occur. We also found that the inferred depths of the convection cells have no noticeable regional dependence. These results suggest that the subsurface of the umbra is an environment where convection can occur more easily than the convergent interface, and hence support the cluster model. For more concrete results, further studies based on umbral velocity oscillations in the lower atmosphere are required.}

   \keywords{sunspots -- Sun: chromosphere -- Sun: oscillations -- Sun: helioseismology
               }
	\titlerunning{Investigation of subsurface structure of a sunspot from umbral flashes}
   \maketitle
%
%-------------------------------------------------------------------

\section{Introduction}

The subsurface structure of sunspots is an important topic in solar physics. This veiled structure situated below the photosphere should be interacting with convective flows, playing a key role in the formation, evolution, and dissipation of the sunspots. The subsurface structure of sunspots is also important in thermal energy transfer. It determines the intensity of the umbrae and gives indication of the formation mechanism of the inhomogeneous small-scale structures inside sunspots \citep{Borrero 2011}. 

Previous studies have proposed two representative models of the sunspot subsurface structure: the monolithic model \citep{Hoyle 1949} and the cluster model \citep{Parker 1979}. The monolithic model treats a sunspot as a single magnetic flux tube. By contrast, in the cluster model, also known as jellyfish or spaghetti model, a sunspot is considered to be an assembly of small flux tubes. There are several observational and theoretical works to verify each model, but they remain inconclusive.

The early magnetic field measurements of sunspots seemed to support the monolithic model. It is well known that a sunspot has a systematic magnetic field structure \citep{Hoyle 1949} that is characterized by monotonically decreasing field strength and increasing field inclination from the sunspot center toward its boundary which favors the monolithic model. However, it is reported that the power of the magnetic field strength oscillations has isolated multiple peaks inside sunspots \citep{Rueedi 1998, Balthasar 1999}. Each penumbral fibril regarded as a magnetic flux tube shows independent behavior, which is referred to as uncombed penumbra \citep{Thomas 2004}. These findings hint towards the cluster model rather than the monolithic model. 

High resolution observations revealed the existence of sub-arcsecond structures inside sunspot umbrae that are represented by umbral dots and light bridges. These structures are closely related to convective motions \citep{Ortiz 2010, Watanabe 2012}. Initially, it was suggested that they cannot be formed in the strong magnetic field environment due to the inhibition of horizontal motions in plasma convection. They were naturally regarded as a tip of the field-free hot gases between magnetic flux tubes in the cluster models \citep{Parker 1979, Choudhuri 1986}. Contrary to this expectation, recent magnetohydrodynamic (MHD) simulations based on the monolithic model successfully reproduced realistic umbral dots \citep{Schussler 2006} or even overall sunspot's fine structures \citep{Rempel 2011}. It implies that the magneto-convection, which is the convection of magnetized plasma, can produce tiny field-free regions inside large flux tubes. Accordingly, it still remains debatable whether these small-scale structures are intrusions between small magnetic flux tubes or internal structures of large flux tubes caused by magneto-convection (See  \citealt{Borrero 2011} and references therein).

The local helioseismology provides the information about the physical quantities below the photosphere. One of these quantities is the subsurface converging flow which is required for the cluster model as it helps individual small flux tubes to keep together. \citet{Zhao 2001} reported a converging downflow 1500 to 5000 km below the photospheric surface of a sunspot. On the contrary, \citet{Gizon 2009} found a horizontal outflow between the solar surface and the depth of 4500 km. Therefore, further helioseismic works need to be carried out for a solid conclusion.

This debate on the sunspot subsurface structure is also inconclusive in theoretical studies. \citet{Meyer 1974} investigated the instability below a sunspot, and conjectured that the small-scale mixing of magnetized and non-magnetized gas will occur below 2000 km which matches a picture of the monolithic model. On the other hand, \citet{Parker 1975} argued that a sunspot below the photosphere cannot be a single magnetic structure due to fluting or interchange instabilities. In conclusion, at present we do not have a solidly established view of the subsurface structure of sunspots. 

To resolve this problem, we investigate here the subsurface structure of a sunspot using umbral flashes. \citet{Beckers 1969b} obtained a time series of  Ca {\sc ii} K-line filtergrams and found that small regions inside sunspot umbrae often brighten for a short period time, and named them umbral flashes. These were later identified with umbral oscillations. \citet{Giovanelli 1972} examined the time series of H$\alpha$ Dopplergrams, and found the oscillations of velocity at fixed points inside sunspot umbrae. It was revealed that these velocity oscillations are upward propagating slow MHD waves \citep{Lites 1984, Centeno 2006}. The umbral flashes are now usually understood as a shock, the nonlinear development of the slow MHD waves \citep{Voort 2003}. 

We expect that umbral flashes provide information about the wave sources in the interior. These sources may be much relevant to the subsurface structure of a sunspot. Recent studies show that it is possible to estimate the wave sources by analyzing oscillatory phenomena \citep{Zhao 2015, Felipe 2017, Kang 2019, Cho 2020}. It is expected that the wave sources are located between 1000 and 5000 km below the photosphere. If we assume that the waves are generated by convective motions, we can infer the distribution of convection cells below the sunspot photosphere. As we mentioned above, umbral dots or light bridges are considered as the photospheric signature of the convection. Previous studies reported the observational evidence for the relation between umbral dots in the photosphere and 3-minute umbral velocity oscillations at the temperature minimum region \citep{Chae 2017, Cho 2019}. It implies that oscillations inside umbrae can be a good tool for investigating the site of the convection below sunspots. 

We determine the spatial distribution of convection cells by analyzing the motion of umbral flashes and infer the subsurface structure of the sunspot based on these results. This is similar to the method of finding faults or plate boundaries of the Earth's crust. Seismic waves of an earthquake typically occur near a convergent boundary of plates and are a useful tool for investigating the internal structure of the Earth. We expect that umbral flashes can be used in a similar way for investigating the source of slow MHD waves and ascertain the subsurface structure of sunspots. 
%--------------------------------------------------------------------
\section{Hypothesis}

An oscillation center is defined as the first detected position of a horizontally propagating oscillation pattern in an umbra \citep{Cho 2019}. Observations support that the wave source is located below the oscillation center. \citet{Cho 2020} nicely demonstrated that the observed horizontal pattern of propagation is very much consistent with the picture of generation of the fast MHD waves by a subsurface source and quasi-spherical propagation in the interior. The convective motions are assumed as the physical mechanism of the wave generation \citep{Lee 1993, Moore 1973}. Therefore, an oscillation center gives us the spatial information about the convection cell below the sunspot. 

It is expected that the wave generation process is different between the two sunspot subsurface structure models. Although it must be magneto-convection that excites waves in both models, the cluster model will generate more waves. The cluster model has field-free region between small flux tubes, whereas the monolithic model does not. The existence of field-free region will facilitate the horizontal motion of plasma, and the wave generation will be easily activated. The main challenge is to determine quantitatively which model can explain the observed wave generation rate in a sunspot.

\begin{figure*}
   \centering
   \includegraphics[width=\hsize]{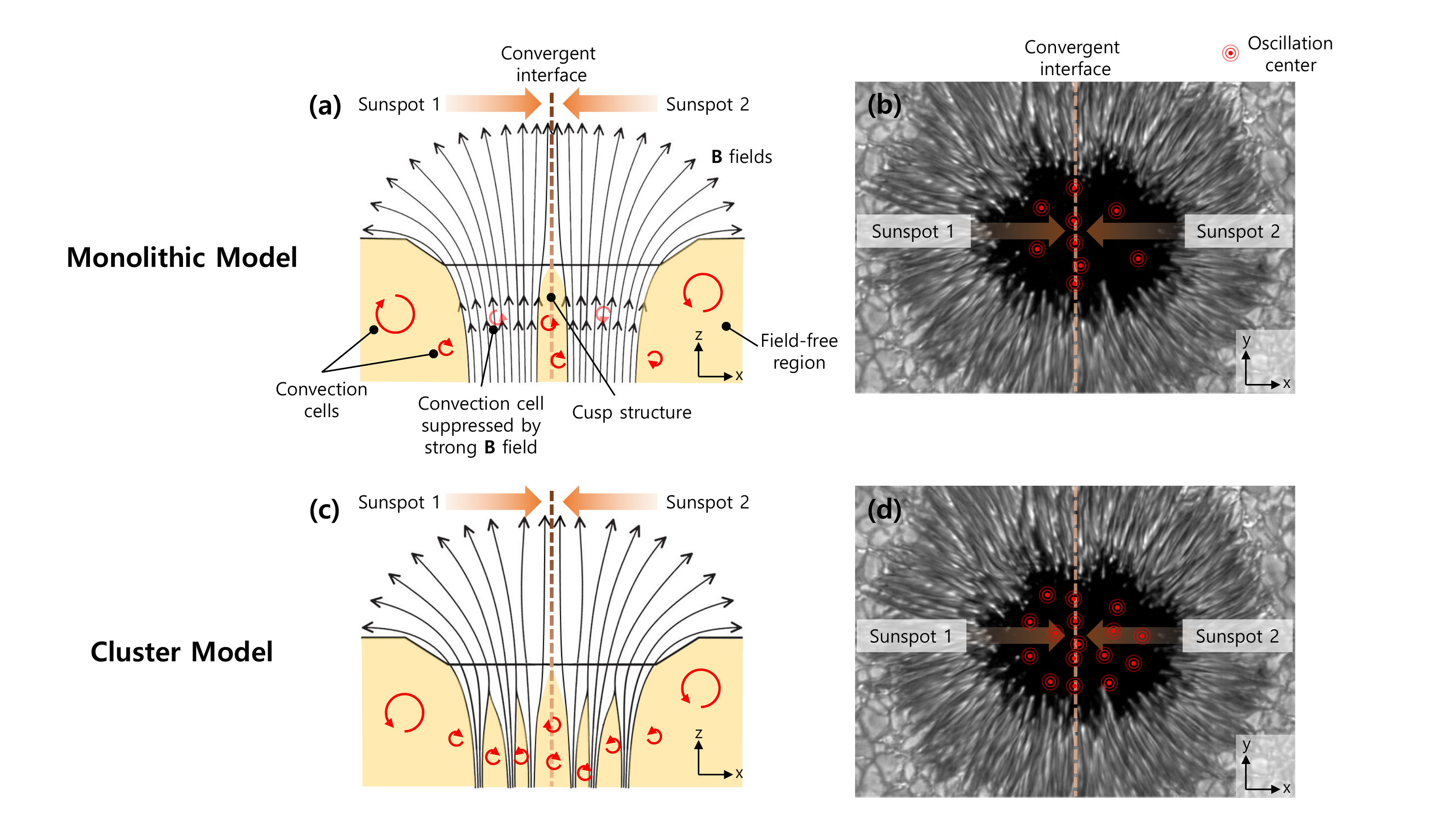}
      \caption{Hypothesis of this study. The left figures show cross sections of two merging sunspots in (a) the monolithic model and (c) the cluster model. The black solid arrows indicate magnetic field lines. The yellow region is a magnetic field-free region. The vertical dashed line indicates the convergent interface. The red circular arrows indicate the general convection cells. The opaque red circular arrows indicate the magneto-convection cells. The right figures (b) and (d) show distributions of the oscillation centers on the plane of the sky. The oscillation centers are marked with red concentric circles.}
         \label{hypothesis}
\end{figure*}

It is, therefore, important to establish a comparison target in order to distinguish between the two models. When two sunspots merge and become one, it is possible to identify the remnant interface between the two sunspots as a light bridge for a while even after merging. We will call this remnant as a convergent interface. This convergent interface is our comparison target. Figure \ref{hypothesis} illustrates our hypothesis. Generally, a sunspot has funnel-shaped field lines. Thus, when two sunspots are merging, a field-free cusp shape structure will be formed on the convergent interface, and is observed as a light bridge in the photosphere. The cusp structure may be maintained for a while even after the end of the merging process at the photosphere. The cusp structures in both models certainly resemble the field-free region in the cluster model. In the monolithic model (Fig. \ref{hypothesis}a), thus, it is expected that more waves will be generated in the cusp structure rather than in the large flux tube. This will cause more oscillation centers to occur near the convergent interface than that on the umbral cores (Fig. \ref{hypothesis}b). If the subsurface structure of the sunspot is close to the cluster model, on the contrary, the cusp structure can be considered as one of the many field-free gaps (Fig. \ref{hypothesis}c). Therefore, the oscillation centers will be found regardless of the convergent interface position (Fig. \ref{hypothesis}d). Hence, the spatial distribution of oscillation centers in relation to the convergent interface serves as a useful observational basis to determine which model is more plausible.

%--------------------------------------------------------------------
\section{Data and Analysis}

For the present study we used data taken by the Interface Region Imaging Spectrograph (IRIS; \citealt{De Pontieu 2014}). We selected data suitable for our study by applying the following criteria. 1) We chose data that captured the merging process of two sunspots. Simple sunspots were favored because their convergent interface can be easy to determine. 2) We selected IRIS silt-jaw images (SJIs) taken through the Mg {\sc ii} 2796 \AA\ filter. The Mg {\sc ii} line is one of the typical chromospheric lines and clearly shows the umbral flashes \citep{Tian 2014, Yurchyshyn 2020}. The SJIs have a spatial sampling of less than 0.33\arcsec\ which is sufficient to identify the shape of umbral flash. 3) A shorter time cadence of 45 s or less is required as the main power of the umbral oscillations is generally three minutes. 4) To obtain large number of successive umbral flashes, we searched for uninterrupted time sequences of about one hour or longer. 

We found that the leading sunspots of AR 12470 are quite suited for our study. The two sunspots initially had a simple round shape, until they merged into a single sunspot. They were observed by IRIS for 3 days from Dec 16 to Dec 18, for about one hour a day. We explored the SJI 2796 \AA\ data for the identification of the umbral flashes, and used simultaneously obtained SJI 2832 \AA\ images to check on the fine photospheric structures inside the umbra. The SJI observations are summarized in Table \ref{SJI data}. Additionally, intensity images and vector magnetograms from the Helioseismic and Magnetic Imager (HMI, \citealt{Schou 2012}) were used as supplementary data. The intensity images show the photospheric long-term evolution of the sunspots, and the vector magnetograms provide information about the temporal evolution of the photospheric magnetic field strength. 
\begin{table*}
\caption{Summary of the IRIS SJI data used in this study}             % title of Table
\label{SJI data}      % is used to refer this table in the text
\centering                          % used for centering table
\begin{tabular}{c c c c c c}       
\hline\hline                 % inserts double horizontal lines
Date \& Time & Duration (min) & Cadence (s) & Location (\arcsec) & Pixel scale (\arcsec) \\    % table heading 
\hline                        % inserts single horizontal line
   2015-12-16  17:11-18:08 UT & 57 & 12.7 & (-480, 223) & 0.33 \\ 
   2015-12-17  19:11-20:06 UT & 55 & 12.8 & (-263, 221) & 0.33 \\ 
   2015-12-18  13:03-14:00 UT & 57 & 12.8 & (-99, 226) & 0.33 \\ 
\hline                                   %inserts single line
\end{tabular}
\end{table*}

\begin{figure*}
   \centering
   \includegraphics[width=\hsize]{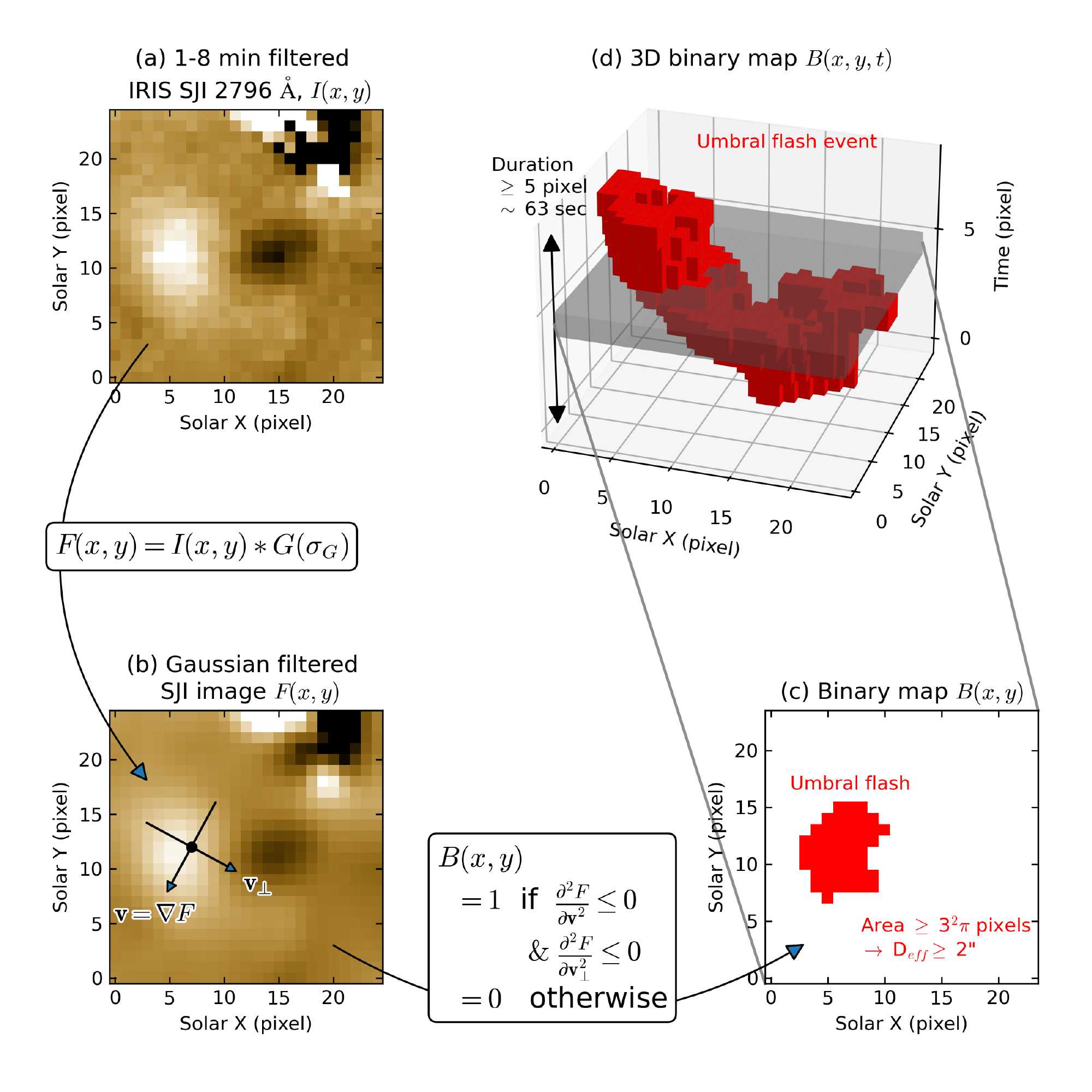}
      \caption{Illustration of how to define a bright patch group. (a) 1--8 min bandpass-filtered IRIS SJI 2796 \AA\ image. (b) Gaussian filtered SJI image. (c) Binary map. The red area indicates the extracted umbral flash. (d) 3D binary map. The red volume indicates an umbral flash event.}
         \label{method}
   \end{figure*}

\begin{figure*}
   \centering
   \includegraphics[width=\hsize]{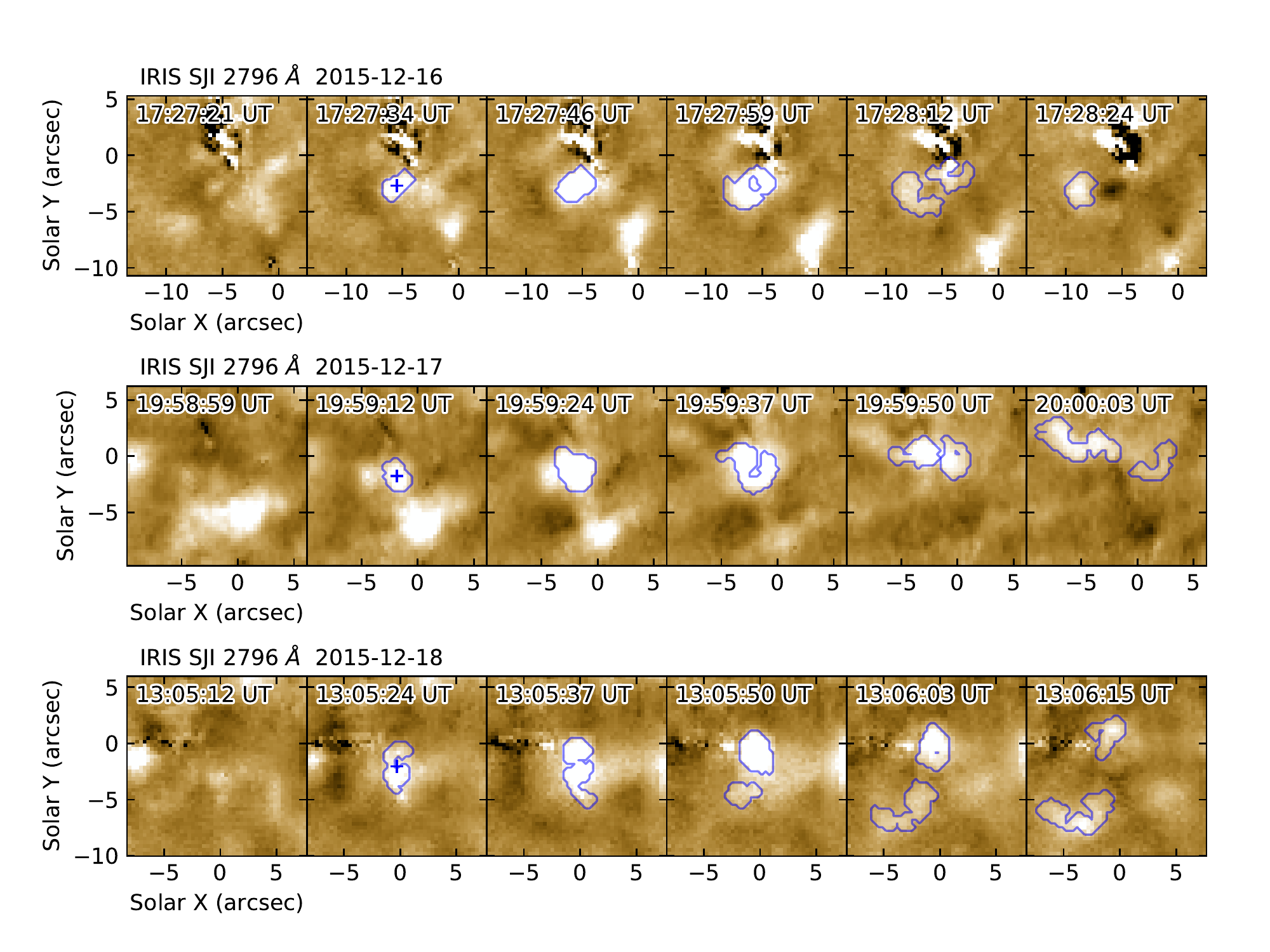}
      \caption{Some examples of the identified umbral flash events. The background image is IRIS SJI 2796 \AA\, and the umbral flashes identified by our method are indicated with blue contours. Each row shows evolution of the identified umbral flashes for different events. The blue cross symbol indicates the oscillation center.}
         \label{method_examples}
   \end{figure*}

Figure \ref{method} illustrates the process of identification of umbral flash events in the IRIS SJI 2796 \AA\ images. First, we applied a 1--8 min bandpass filter to the data $I(x,y)$ to reduce the noise and other instrumental effects (Fig. \ref{method}a). One can find an umbral flash that appears to be a bright patch in that image. We applied a 2D Gaussian filter $G(\sigma_G)$ with a width $\sigma_G$ of 1 pixel to simplify the shape of the umbral flashes (Fig. \ref{method}b). We then extracted the area of the umbral flashes. The umbral area was defined by an ellipse with major and minor axes of about 23\arcsec and 13\arcsec, respectively. We calculated the second derivative sign in the gradient direction and the perpendicular to the gradient direction for every pixel and generated a binary map $B(x, y)$ by collecting pixels which have a negative value of the both second gradient signs (Fig. \ref{method}c). We defined spatially connected pixels that have an area $S$ greater than $9\pi$ pixels as the umbral flash. It means that we only selected the umbral flashes with effective sizes ($2\sqrt{S/\pi}$) greater than 2\arcsec.

We produced a 3D binary cube $B(x, y, t)$ by stacking binary maps for each observing time. In the binary cube, a group of connected voxels can be obtained (Fig. \ref{method}d). We defined the group of voxels as an umbral flash event, which is regarded as a phenomenon that occurs as a wavefront with the same propagation phase \citep{Cho 2020}. We only selected umbral flash events with a lifetime longer than 65 seconds (identified in 5 successive images). The effective size of the umbral flash events was defined as the mean value of the effective size of the umbral flashes composing that event. The time cadence of the IRIS SJIs is short enough to follow individual umbral flash events. In our data, the typical size and horizontal speed of the umbral flashes are approximately 2000 km and 40 km s$^{-1}$, respectively (see Figs 4 \& 7b), which is consistent with previous reports (e.g. \citet{Beckers 1969}). It implies that an umbral flash generally takes about 50 s to move away from its original position, which is much longer than the IRIS SJIs' time cadence of 13 s. There might be a concern that extremely fast-moving umbral flashes (2000 km / 13 s $\gtrsim $  150 km s$^{-1}$) could be identified as separate umbral flash events. However, such an event will be split into several events with shorter lifetimes, so it is very likely that they are filtered out by our lower limit of lifetime (65 s).

The center of each umbral flash UF$_{\textrm{cen}}$ was determined by the 2796 \AA\ intensity-weighted center 
\begin{equation}
{\textrm{UF}}_{\textrm{cen}}(\mathbf{r}) = \frac{\sum_i I_{2796, i} \mathbf{r}_i}{\sum_i I_{2796, i}} .
\end{equation}
The position of an UF$_{\textrm{cen}}$ is located inside that umbral flash except in some special cases when the umbral flashes are in the shape of an arc or concentric circle. These kinds of cases, however, are rare, so we think the resulting error is statistically insignificant. We analyzed the motion of umbral flashes using the identified centers UF$_{\textrm{cen}}$. The center of the first detected umbral flash, which is the bottommost part in the 3D voxels, was determined as the oscillation center of the umbral flash event. This method is similar to that of previous studies using umbral flashes \citep{Liang 2011, Yurchyshyn 2020} and umbral velocity oscillations \citep{Cho 2019}. In these studies, their origins were manually identified as the first detected positions of the umbral flashes or umbral oscillations. Figure \ref{method_examples} shows some examples of umbral flash events detected by our method. The merging or splitting of umbral flash events was frequently detected. We counted the umbral flash events based on the oscillation centers. We expect that the merging of different events may cause a small error in tracking their position.

%--------------------------------------------------------------------
\section{Results}

\begin{figure}
   \centering
   \includegraphics[width=\hsize]{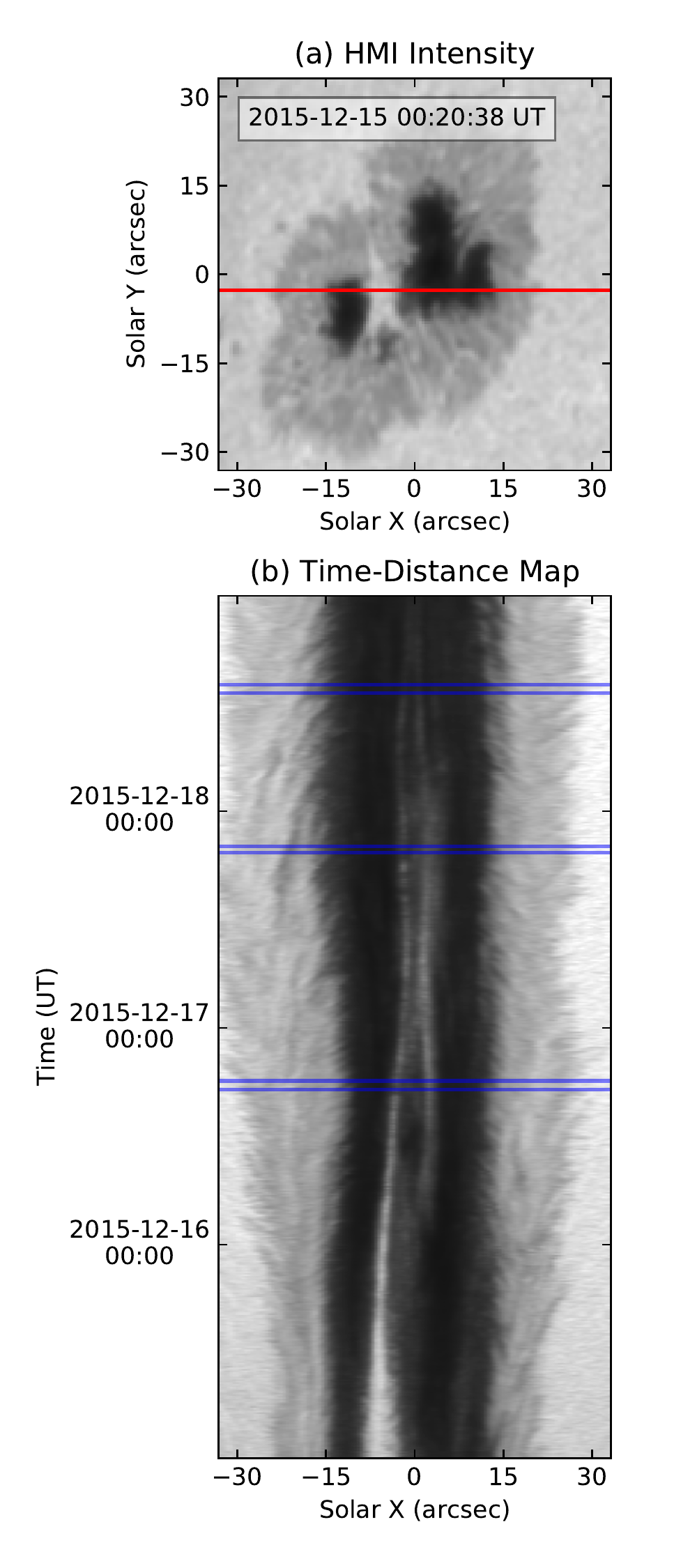}
      \caption{(a) Initial stage of the merging sunspot. This photospheric observation was performed by SDO/HMI intensitygram on Dec 15, 2015. The red solid line indicates slit position for the following time-distance map. (b) Time-distance map of intensity from Dec 15 to 18, 2015. The three time domains marked with blue solid lines indicate the time covered by IRIS SJI data sets.}
         \label{evolution}
   \end{figure}

Figure \ref{evolution} and an animation associated with Fig. \ref{evolution ani} show the evolution process of the merging sunspot. It is very likely that there were two separated magnetic field structures at the initial stage. The two primordial sunspots were divided by a strong light bridge (Fig. \ref{evolution}a). The strong light bridge had a width of about 7\arcsec\ and appears as a narrow photospheric region similar to the region outside of the sunspots. The light bridge became thinner and weaker with time, then eventually it disappeared (Fig. \ref{evolution}b). Finally, the two sunspots merged and became a single round sunspot. This process occurred within three days. The three SJI data sets we found show three stages of the light bridge: the thin, faint, and extinct stage. The light bridge dividing the two sunspots existed near x=0\arcsec at Dec 16, so we assumed an imaginary line with x=0\arcsec as the convergent interface.

\begin{figure*}
   \centering
   \includegraphics[width=\hsize]{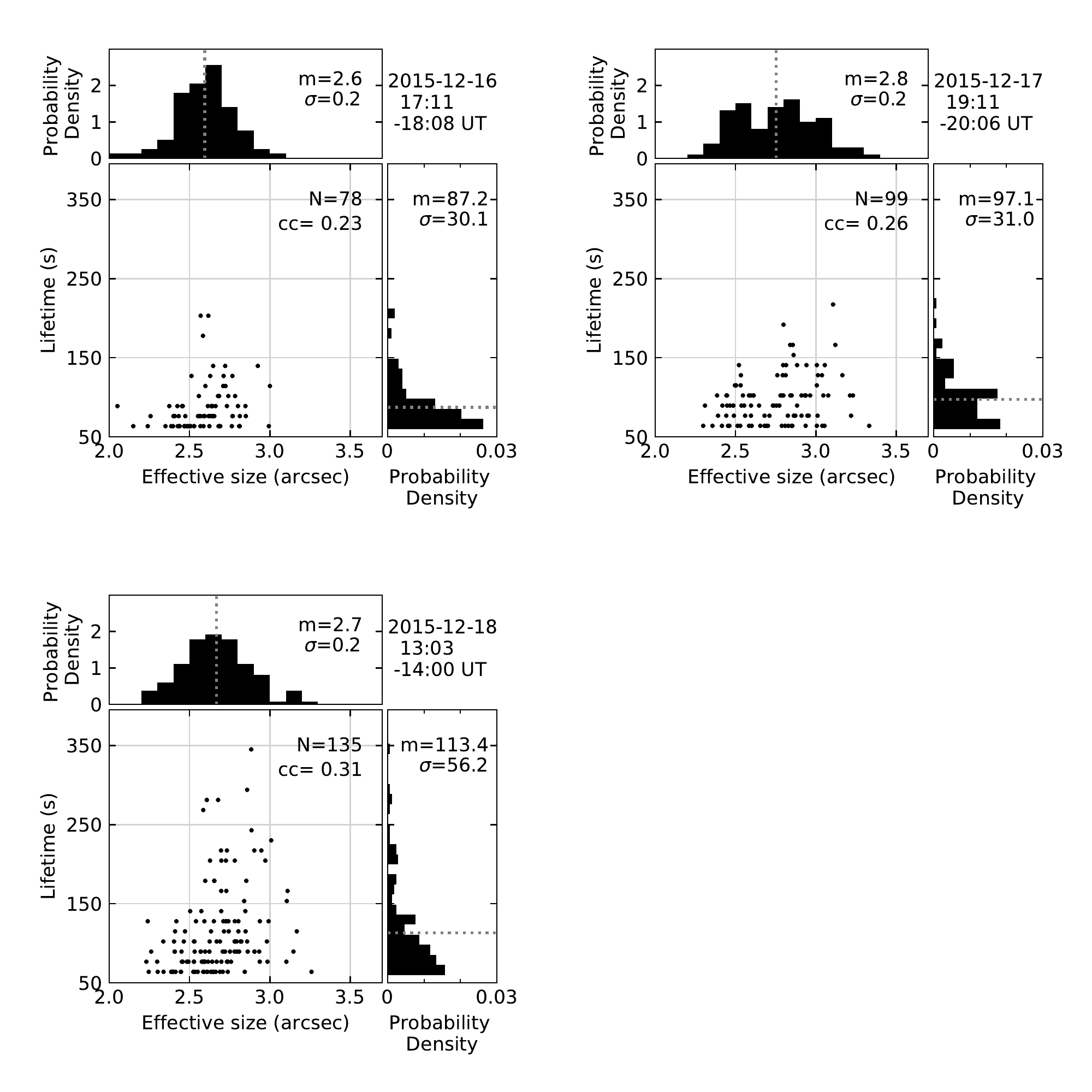}
      \caption{Scatter plots between the effective size and lifetime of identified brightening features from the three SJI 2796 \AA\ data sets. The effective size and lifetime histograms are shown in the upper and in the right panel, respectively. The date, time, total number of the identified umbral flash events $N$, mean value $m$, and standard deviation $\sigma$ are indicated in each panel. The dotted gray line indicates the mean value of each parameter. Note that 2\arcsec\ and 65 s are the methodological lower limits of the effective size and lifetime, respectively.}
         \label{stat1}
   \end{figure*}

We identified about one thousand umbral flashes for each of the SJI 2796 \AA\ data sets. Only some of them satisfied our criterion described in the previous section and were counted as umbral flash events. The total number of selected umbral flash events increased from 78 to 99 and 135 for the three data sets. 

Figure \ref{stat1} shows the distributions of two parameters, the effective size and lifetime, of the selected umbral flash events. The effective size shows a normal distribution with a mean value of about 2.7\arcsec\ and a standard deviation of about 0.2\arcsec. These values did not change significantly over time. Concerning the lifetime, shorter lifetime events were dominant. Most of the umbral flash events had a lifetime shorter than 150 seconds. The proportion of long-lasting events slightly increased with time. The effective size and the lifetime had a weak positive correlation. Long-lasting umbral flash events consisted of relatively larger umbral flashes. It is not clear whether this tendency reflects the nature of the umbral flash events, or it is caused by the limitation of the algorithm related to the merging of the umbral flash events. 

\begin{figure*}
   \centering
   \includegraphics[width=\hsize]{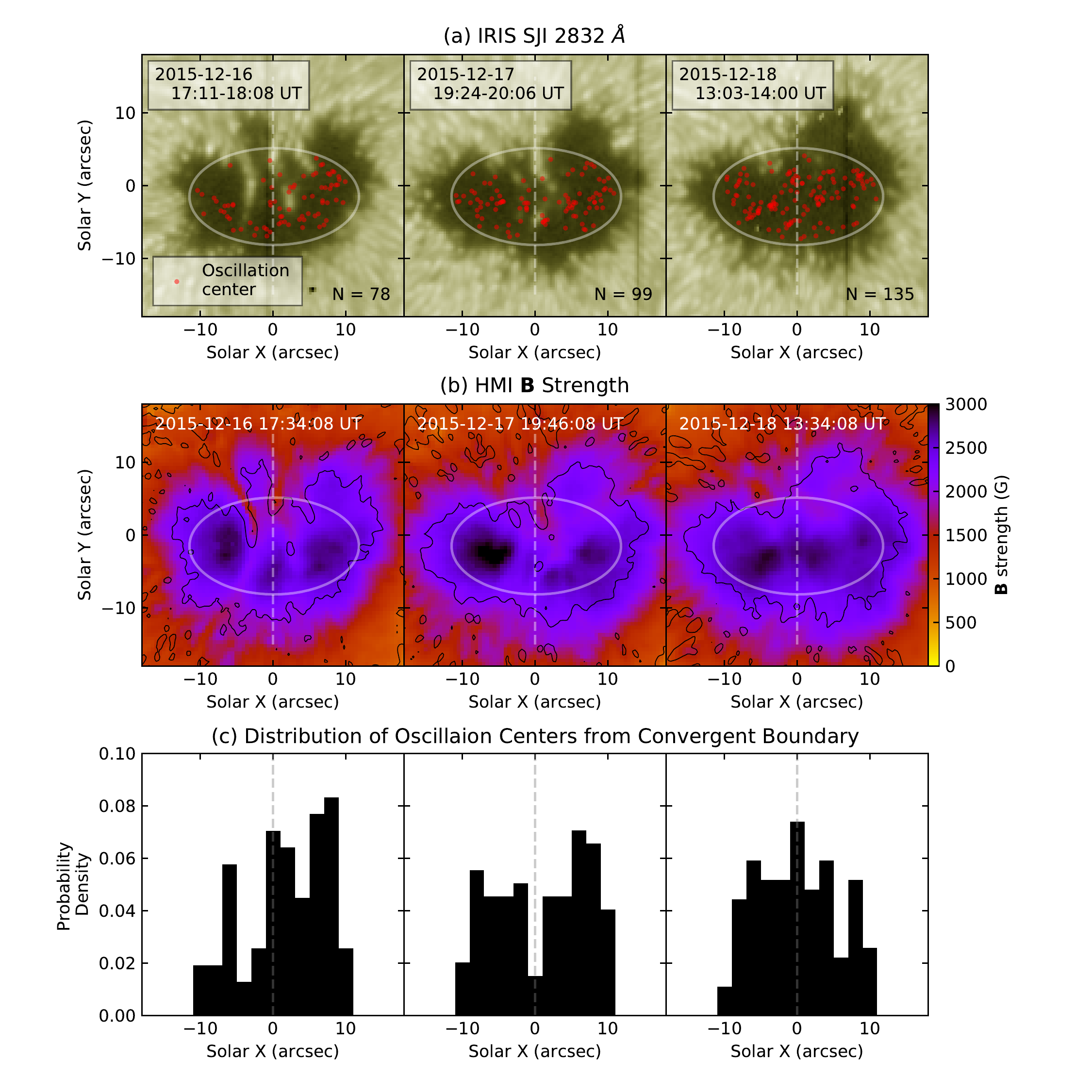}
      \caption{(a) Spatial distribution of the oscillation centers (red dots). The background image is the IRIS SJI 2832 \AA\ that shows the solar atmosphere at the photospheric height. The white ellipse and dashed line indicate a defined umbral region for identifying the umbral flashes and the convergent interface, respectively. (b) HMI magnetic field strength map. The black contour indicates the umbral-penumbral boundary obtained from the IRIS SJI 2832 \AA\ image. (c) Distribution of the oscillation centers from the convergent interface. Each column represents a different observation date.}
         \label{oc_map}
   \end{figure*}

Figure \ref{oc_map}a and \ref{oc_map}b show the detailed structure of the sunspot. On the first day, the light bridge was split into two strands near the convergent interface. The two strands were deformed, weakened, and slightly closer to each other on the second day, and finally disappeared on the third day. The light bridge had weak magnetic field strength that was about 1000 G weaker than that of the umbral core which is consistent with previous reports \citep{Beckers 1969, Rueedi 1995, Jurcak 2006}.

Figure \ref{oc_map}a shows the spatial distribution of the oscillation centers for each data set. The most important finding is that the oscillation centers appear without any noticeable spatial relationship to the convergent interface. They are not crowded near the convergent interface marked with the vertical dashed line in Fig. \ref{oc_map}a. Figure \ref{oc_map}c also clearly shows that the convergent interface (x=0\arcsec) is not a special region for the oscillation centers. This characteristic persists in all three data sets.

A careful examination reveals some details on the spatial distribution of the oscillations. We find that there are voids, the areas where oscillation centers are either absent or very few. These voids may be partly attributed to the difficulty of identifying umbral flashes. The most representative example of such a void is the area around (-4", 0") on the first day (see Fig \ref{oc_map}a). There was a conspicuous light bridge in the photosphere of this area, and we found just a few oscillation centers within a radius of about 3\arcsec, which is matched the abscissa value x=$-$4\arcsec\ in Fig. \ref{oc_map}c, first panel. We identified recursive strong jets anchored in this light bridge from the IRIS SJI 2796 \AA\ data. This kind of jet occurrence may have kept us from clearly identifying the umbral flashes. Another example is found on the third day. In the right column of Fig. \ref{oc_map}c, one can find a dip at x=7\arcsec. This is because the IRIS slit, shown as a dark vertical line in the right column of Fig. \ref{oc_map}a, had a fixed position as the observations were taken in sit-and-stare mode on the third day. Thus, the umbral flashes were not properly identified at there on the third day.

We found that the voids are results of the identification problem of umbral flashes, but it seems that some voids located on bright photospheric regions such as light bridges and umbral dots, are real. In Fig. \ref{oc_map}a, it can be seen that the oscillation centers are hardly found at the core of the bright umbral dots or the light bridges, but several oscillation centers are located at their edges. Figure \ref{oc_map}c shows this fact more clearly. In the histograms, dips are found at x=4\arcsec\ on the first day and x=0\arcsec\ on the second day. It indicates that fewer oscillation centers are found in the locations where bright umbral dots and light bridges were concentrated. We could not find strong jet phenomena there.

\begin{figure*}
   \centering
   \includegraphics[width=\hsize]{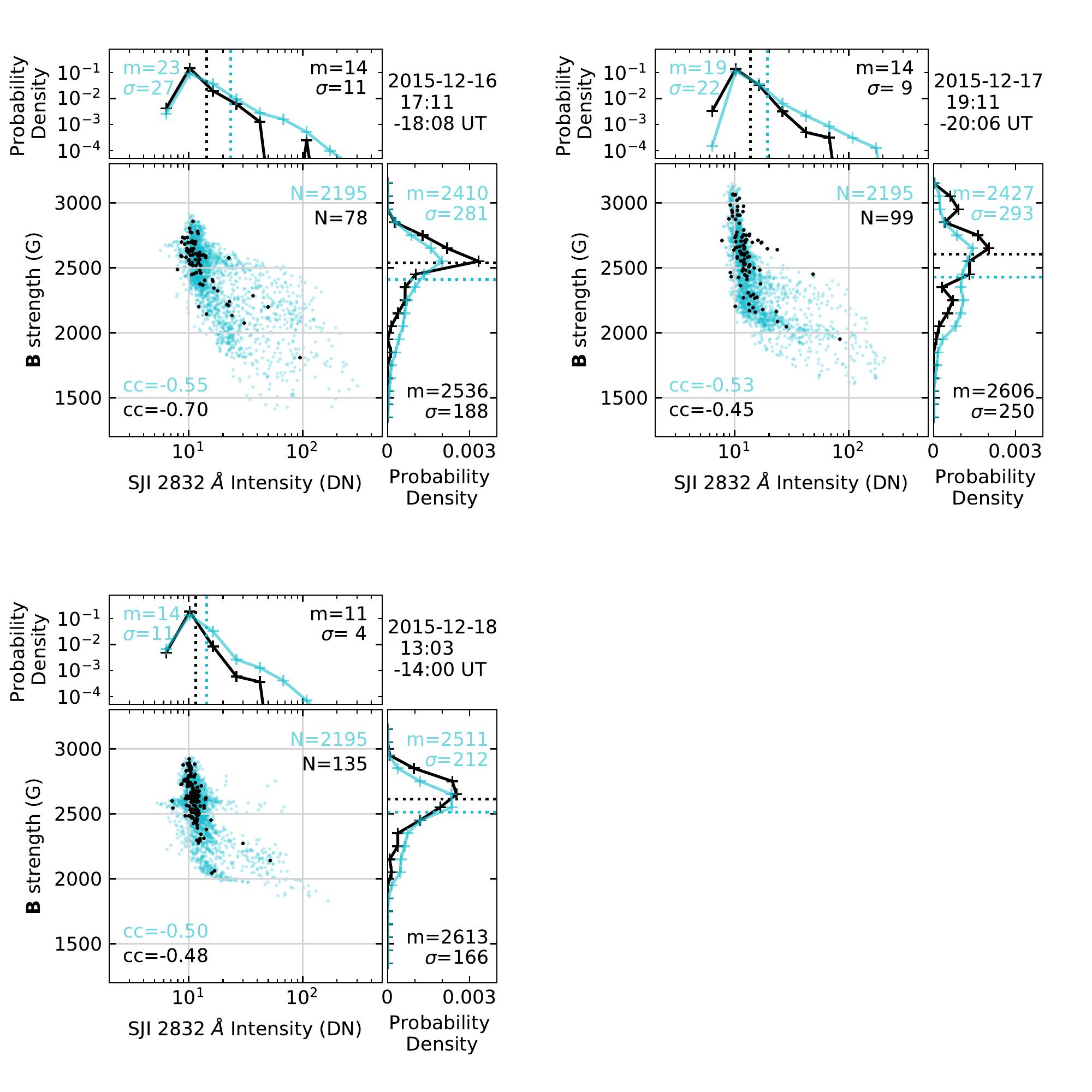}
      \caption{Scatter plots between the IRIS SJI 2832 \AA\ intensity and the HMI magnetic field strength. The format is similar to Fig. \ref{stat1}. The IRIS SJI 2832 \AA\ intensity and magnetic field strength histograms are shown in the upper and the right panel of each scatter plot, respectively. The dark and blue plots indicate the results from the oscillation centers and the whole umbra defined by the inside of the ellipse in Fig. \ref{oc_map}a, respectively. Note that the IRIS SJI 2832 \AA\ intensity is plotted in a logarithmic scale.}
         \label{stat2}
   \end{figure*}

We investigated the characteristics of the IRIS SJI 2832 \AA\ intensity and magnetic field strength in the HMI data at the oscillation centers (Fig. \ref{stat2}). The black scatterplot shows the relation between the two parameters at the oscillation centers. We also plotted the results from the IRIS SJI pixels composing the umbra in blue color. Both scatterplots clearly exhibit a negative correlation. However, there are differences between the photospheric characteristics at the positions of the oscillation centers and those in the whole umbra. The oscillation centers tend to occur in darker photospheric regions and stronger magnetic field locations known as dark nuclei of the umbral core \citep{Sobotka 1993}. This is more clearly seen in each histogram. The probability density function of the photospheric intensity at the oscillation centers is similar to those from the whole umbra but slightly more inclined toward lower intensity. The mean values of the IRIS SJI 2832 \AA\ intensities at the oscillation center positions range from 11 DN to 14 DN, which are about 0.5 $\sigma$ lower than those from the whole umbra. Similarly, the oscillation centers are located above regions with a relatively stronger magnetic field strength. The mean values of magnetic field strength at the oscillation centers range from about 2500 G to 2600 G, which is about 0.5 $\sigma$ higher than those of the whole umbra in each dataset. These results agree with our finding in the above that the oscillation centers are not found near the light bridge and umbral dots (see Fig. \ref{oc_map}).

\begin{figure*}
   \centering
   \includegraphics[width=\hsize]{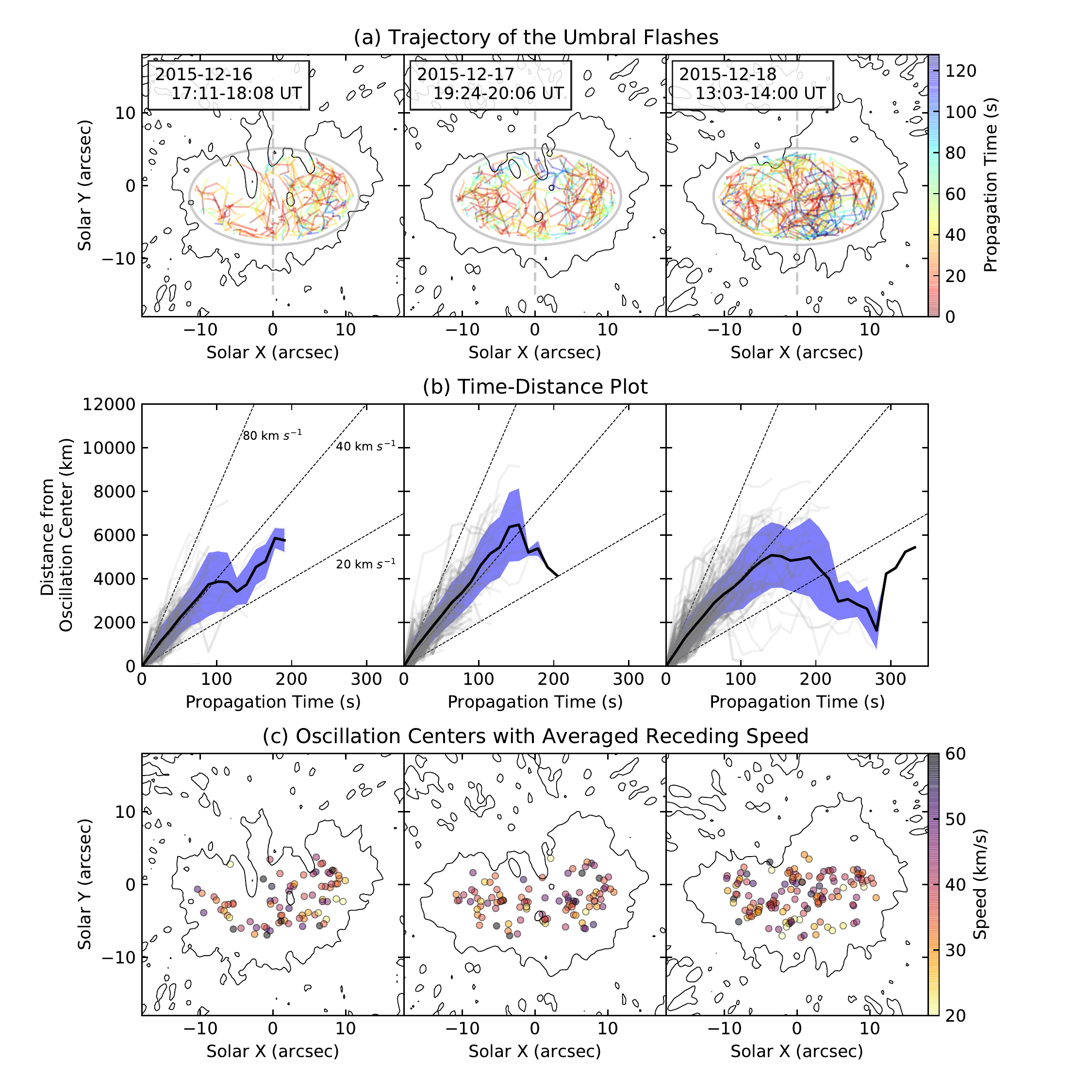}
      \caption{(a) Trajectory of the identified umbral flashes. The solid line indicates the movement of the center of the umbral flashes for each event. The color indicates the propagation time from the beginning of each event. (b) Time-distance plot of umbral flash motion. Each gray solid line represents the propagation of umbral flash from its oscillation center. The black solid line and blue area indicate the mean value of the distance and standard deviation of the distance. The guidelines at 20, 40, and 80 km s$^{-1}$ are shown with dashed lines. (c) Positions of the oscillation centers with averaged receding speed. The color corresponds to the averaged radial propagation speed, which is equivalent to the mean gradient of each gray line in (b). The black contour indicates the umbral-penumbral boundary in the IRIS SJI 2832 \AA. Each column represents a different observational date.}
	\label{traj}
\end{figure*}

We also analyzed the motions of the umbral flashes (Fig. \ref{traj}a). The behavior of motion in each umbral flash is quite complex, displaying azimuthal or rotating motion as well, which is consistent with previous reports \citep{Voort 2003, Sych 2014, Su 2016, Felipe 2019, Kang 2019}. Their behavior is quite distinct from the systematic photospheric motions of umbral dots and penumbral grains that move radially towards the center of the umbra \citep{Sobotka 2009, Kilcik 2020}. 

We find that umbral flashes recede from their oscillation centers. Figure \ref{traj}b clearly shows that the distance between umbral flashes and their oscillation center increased with time in most cases. The receding speed ranges from 20 km s$^{-1}$ to 80 km s$^{-1}$ with a mean value of about 40 km s$^{-1}$. This mean value persisted for about 150 s, and after that it started to decrease. We speculate that this decrease of the receding speed is presumably an error caused by the merging of long-lasting umbral flashes. Our results on the receding motion of umbral flashes are similar to the results on the apparent horizontal propagation of umbral oscillations previously reported by \citet{Cho 2019}, \citet{Kang 2019}, and modelled by \citet{Cho 2020}.

We are able to estimate the source depth of the waves from the trajectory of the umbral flashes. \citet{Cho 2020} described the radial propagation of simple concentric umbral oscillations as an apparent motion. They modeled quasi-spherically propagating fast MHD waves below the photosphere and showed that the time lag near the photospheric level causes the apparent motion. If so, the radially receding speed from the oscillation center should be closely related to the source depth. A faster radially receding speed corresponds to a deeper source depth, which was also demonstrated by \citet{Felipe 2017} using numerical simulations. This relation is not only limited to a simple concentric propagation. \citet{Kang 2019} presented an analytic solution of spiral wave patterns which appear as rotating umbral oscillations patterns with very complicated shapes. They showed that the radial and azimuthal motions are caused by fast MHD waves originated from a 1600 km source depth and cylindrical harmonic oscillations, respectively. Therefore, we believe that the radially receding speed shown in Fig. \ref{traj}b and \ref{traj}c is also closely related to the source depth although the model in \citet{Cho 2020} is simple, and the trajectory of the umbral flashes found here were chaotic (Fig. \ref{traj}a). 

According to this interpretation, our results imply that the depth of the wave sources also seems not to be associated with the convergent interface. \citet{Cho 2020} showed that an average receding speed of 20 km s$^{-1}$ to 40 km s$^{-1}$ indicates about a 2000 km to 5000 km source depth. Simply applying this relation to our results, bright and dark dots in Fig. \ref{traj}c indicate shallow ($\sim$2000 km) and larger ($\geq$5000 km) depths, respectively. We can therefore conclude that shallow and deep sources were mixed inside the umbra, and there was no regional dependency of the source depth. 

%--------------------------------------------------------------------
\section{Discussion}

In this study, we investigated the subsurface structure of a merging sunspot by analyzing umbral flashes extracted from IRIS SJI Mg {\sc ii} 2796 \AA\ images. We developed an algorithm to identify the umbral flashes and umbral flash events. Most of the umbral flash events had a lifetime shorter than 150 s, and an effective size of about 2.7\arcsec. By tracking the motion of the umbral flashes, we were able to determine the position of the oscillation centers. They are distributed over the whole umbra regardless of the convergent interface. They preferentially exist at the dark nuclei and tend to avoid the regions such as light bridges and umbral dots. We also estimated the source depth of the waves. According to our analysis, the source depths showed no clear spatial dependence, either. These characteristics related to the umbral flashes little changed during the three merging stages. 

Our finding of the distribution of the oscillation centers is consistent with previous reports. \citet{Liang 2011} determined the positions of wave sources in Ca {\sc ii} H intensity image series, which is similar to oscillation centers. The identified 19 sources were found near the umbral cores away from the light bridge. \citet{Su 2016} studied the interference of umbral waves on light bridges. They found that the umbral waves do not propagate from the light bridge, but move toward the light bridges. This also implies that oscillation centers do not exist near the light bridges. \citet{Yurchyshyn 2020} reported that the origin of umbral flashes is mainly found in dark nuclei with a magnetic field strength of about 3000 G. \citet{Voort 2003} and \citet{Sych 2020} also found that sources of chromospheric waves are displaced relative to the photospheric umbral dots. In line with these studies, we obtained the results that the oscillation centers are mainly located at the dark nuclei rather than near bright and weak field regions.

Our results support the notion that the subsurface structure of the sunspot is close to the cluster model. The oscillation centers are spread over the whole umbra. It implies that convection cells as the wave sources also exist everywhere beneath the photosphere. We emphasize that what we investigated is not a typical single sunspot but a merging sunspot. Even though the merging sunspot consisted of two distinct environments represented by umbral cores and a convergent interface, it seems that this kind of special configuration does not affect the position of the convection cells. It signifies that the convergent interface of the merging sunspot may not differ from the umbral cores in the aspect of the wave generation. In other words, the field-free gaps that are supposed to exist below the convergent interface, are likely to exist even below the subregions in the umbral cores. This finding matches well with the cluster model.

The distribution of the source depth also supports the cluster model. There was no regional dependency on the source depth. It indicates that the environment of the wave generation was not much different between the convergent interface and the umbral core. Moreover, all the characteristics from analyzing the umbral flashes persisted for the three days. This fact may imply that the subsurface structure of the sunspot did not change dramatically during the merging process, because the sunspot was already an aggregation of small flux tubes. Therefore, we conjecture that the subsurface structure of the sunspot was similar to the cluster model.

Interestingly, our results are contrary to our previous studies from velocity oscillations. \citet{Chae 2017} found strong oscillation power near a light bridge and umbral dot groups. Similarly, \citet{Cho 2019} reported that individual umbral oscillation patterns originate from umbral dots. These findings of our previous studies are incompatible with our current results. This discrepancy may be partially resolved by noting that both previous studies used velocity oscillations inferred from Fe {\sc i} 5435 \AA\ line, which is formed around the temperature minimum region. There are two aspects we have to consider. One is the position error of oscillation centers caused by the propagating direction of waves from the low atmospheric level to the higher level. The other is the efficiency of nonlinearly developing strong intensity fluctuations from waves.

First, we note that the horizontal locations of oscillation centers may change with height if weaves cannot propagate predominantly in the vertical direction. It is reported that the magnetic field configuration over light bridges is found to be complex \citep{Felipe 2016, Yuan 2016, Toriumi 2015}. Similarly, umbral dots have a higher magnetic field inclination than other umbral regions \citep{Socas-Navarro 2004, Watanabe 2009}. Accordingly, as slow MHD waves propagate along the non-vertical magnetic field, the oscillation center determined at the upper atmosphere may deviate from the horizontal position of the wave source in a light bridge or an umbral dot. 

Second, we note that umbral flashes favor to occur in strong magnetic field environments. The intensity oscillations such as umbral flashes are considered as a result of highly compressed gas by nonlinear propagation of slow MHD waves. It is likely that the gas compression may not fully develop in relatively weak field regions because the waveguiding effect of magnetic field is expected to be weaker in weaker field regions than in stronger field regions. This will cause a lower occurrence rate of umbral flashes in weak field medium. Light bridges are relatively weak field regions at the photosphere \citep{Beckers 1969, Rueedi 1995, Jurcak 2006}. If this weak field environment extends to the upper atmosphere, fewer umbral flashes will be found above the light bridges. Umbral dots also are not free from this effect because of their weak field nature \citep{Socas-Navarro 2004, Riethmuller 2008}. Therefore, the configuration and strength of the magnetic field may cause a biased detection of wave phenomena.  

To fully resolve the above discrepancy, further research is required based on velocity oscillations in the lower atmosphere. The velocity is more useful than intensity for the description of waves in a medium. Measurements at lower heights reduce the error of the oscillation center position arose from the magnetic field inclination. Therefore, it is required to identify the relation between velocity oscillations at lower atmosphere and chromospheric umbral flashes and to draw a more solid conclusion on the subsurface structure of sunspots using it.

\begin{acknowledgements}
We greatly appreciate the referee's constructive comments. This research was supported by Basic Science Research Program through the National Research Foundation of Korea (NRF) funded by the Ministry of Education (NRF-2020R1I1A1A01068789). MM acknowledges DFG-grant WI 3211/8-1. MM was supported by the National Research Foundation of Korea (NRF-2019H1D3A2A01099143, NRF-2020R1A2C2004616). IRIS is a NASA small explorer mission developed and operated by LMSAL with mission operations executed at NASA Ames Research Center and major contributions to downlink communications funded by ESA and the Norwegian Space Centre.
\end{acknowledgements}


\begin{appendix} %First appendix
\section{Online material}
   
   \begin{figure}
   \centering
   \includegraphics[width=\hsize]{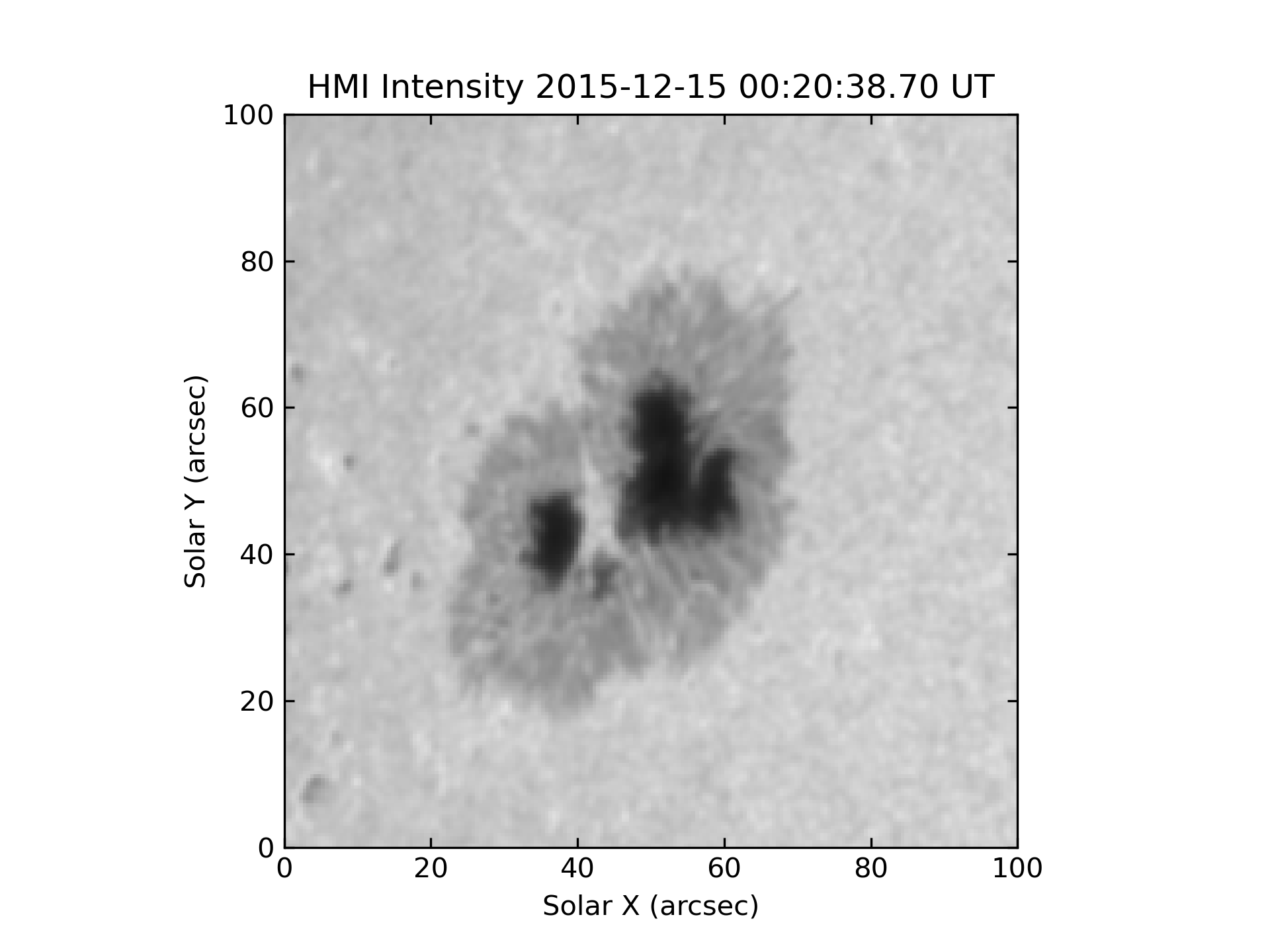}
      \caption{Animation sequence of the HMI intensitygram. It shows sunspot evolution for 3 days from 2015 Dec 15 00:20 UT.}
         \label{evolution ani}
   \end{figure}

\end{appendix}

\end{document}